\documentclass[prl,twocolumn,amsmath,amssymb,aps,superscriptaddress,preprintnumbers]{revtex4-2}

\usepackage{graphicx}
\usepackage{braket,slashed,changepage,placeins,multirow,colortbl,makecell,amsmath,mathtools,array,physics}
\usepackage{xspace}
\usepackage{enumitem}
\usepackage[T1]{fontenc}
\usepackage[utf8]{inputenc}
\usepackage{float}  
\usepackage[colorlinks=true,citecolor=blue,urlcolor=blue,linkcolor=blue,anchorcolor=blue]{hyperref}
\usepackage{bbold}
\newcommand{\fat}[1]{\boldsymbol{#1}}

\newcommand{\cB}{\ensuremath{\mathcal{B}}}
\newcommand{\cC}{\ensuremath{\mathcal{C}}}

\newcommand{\LQCD}{\ensuremath{\Lambda_{\rm QCD}}}

\newcommand{\nbar}{\bar{n}} 

\newcommand{\bO}{\ensuremath{\mathbf{O}}}
\newcommand{\cO}{\ensuremath{\mathcal{O}}}
\newcommand{\pbar}{\bar{p}}

\newcommand{\cP}{\ensuremath{\mathcal{P}}}

\newcommand{\cU}{\ensuremath{\mathcal{U}}}

\newcommand{\ddslash}{{d\!\!{}^-}}
\newcommand{\IInt}{\int\!\!\!\!\!\int}
\newcommand{\deltaslash}{{\delta\!\!\!{}^-}}
\newcommand{\collapsel}{\Big\{\!\!\!\Big\{}
\newcommand{\collapser}{\Big\}\!\!\!\Big\}}
\newcommand{\scollapsel}{\{\!\!\{}
\newcommand{\scollapser}{\}\!\!\}}

\DeclareMathOperator*{\sumint}{%
\mathchoice%
{\ooalign{$\displaystyle\sum$\cr\hidewidth$\displaystyle\int$\hidewidth\cr}}
{\ooalign{\raisebox{.14\height}{\scalebox{.7}{$\textstyle\sum$}}\cr\hidewidth$\textstyle\int$\hidewidth\cr}}
{\ooalign{\raisebox{.2\height}{\scalebox{.6}{$\scriptstyle\sum$}}\cr$\scriptstyle\int$\cr}}
{\ooalign{\raisebox{.2\height}{\scalebox{.6}{$\scriptstyle\sum$}}\cr$\scriptstyle\int$\cr}}
}

\def\nn{{\nonumber}}

\begin{document}

\preprint{MIT-CTP 6060, LA-UR-26-20114}

\title{Factorization of elastic, single, and double diffractive $pp$ scattering}

\author{Philipp B.~Aretz}
\email{aretz24@mit.edu}
\affiliation{MIT Center For Theoretical Physics -- A Leinweber Institute, Cambridge, MA 02139, USA}

\author{Kyle Lee}
\email{kyle@anl.gov}
\affiliation{High Energy Physics Division, Argonne National Laboratory, Lemont, IL 60439, USA}

\author{Stella T.~Schindler}
\email{schindler@lanl.gov}
\affiliation{Physics and Theoretical Divisions, Los Alamos National Laboratory, Los Alamos, NM 87545, USA}

\author{Iain W.~Stewart}
\email{iains@mit.edu}
\affiliation{MIT Center For Theoretical Physics -- A Leinweber Institute, Cambridge, MA 02139, USA}

\begin{abstract} 
We use effective field theory techniques to factorize elastic, single, and double diffractive forward $pp$ scattering in the Regge limit $|t|\ll s$, where $t$ is the squared momentum transfer. These processes involve a large rapidity gap and comprise about half the total $pp$ cross section.
We explain why the diffractive PDFs appearing in $ep$ diffraction do not appear as universal hadronic functions for $pp$ diffraction.
For $|t|\sim \Lambda_{\rm QCD}^2$, we show that the hadronic functions in $ep$ and $pp$ diffraction differ, and hence are non-universal.
In general, we prove that rapidity anomalous dimensions are universal between diffractive $ep$ and $pp$ processes, and that color-singlet (Pomeron) evolution equations can be determined at the amplitude level. 
\end{abstract}

\maketitle

Diffraction encompasses nearly half the cross section at the Tevatron and LHC \cite{Deile:2006tt, Baltz:2007kq, Feng:2022inv, LHCForwardPhysicsWorkingGroup:2016ote, ParticleDataGroup:2024cfk}. In $pp$ diffraction, two protons forward-scatter off one another, producing final states exhibiting a large rapidity gap \cite{Feinberg1956, Chew:1961ev}. 
Deciphering diffraction is critical to understanding the total cross section \cite{CDF:1993wpv, TOTEM:1999aa, TOTEM:2013lle, ATLAS:2022mgx}, building accurate Monte Carlo generators \cite{Bierlich:2022pfr, Rasmussen:2018dgo, Gieseke:2016fpz, Bellm:2017bvx, Sherpa:2024mfk}, and probing wide-ranging phenomena, from the structure of nuclear matter and its saturation with gluons \cite{Newman:2013ada, AbdulKhalek:2021gbh, Hentschinski:2022xnd, Frankfurt:2022jns} to the behavior of cosmic ray cascades \cite{Knapp:2002vs,Grieder2010, Engel:2011zzb}. Here, we derive the QCD factorization of elastic, single, and double $pp$ diffraction in terms of first-principles field theory operators.

\begin{figure}[t]
	\includegraphics[width = 2 in]{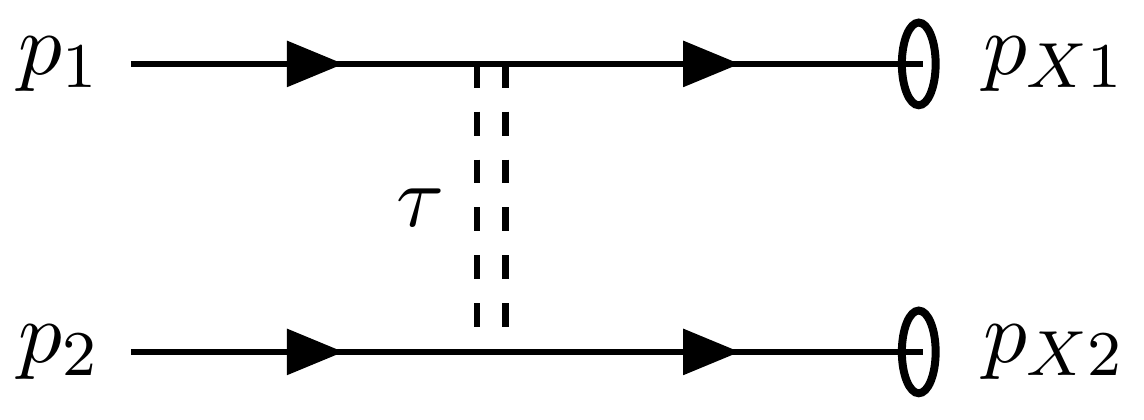}
	\caption{Elastic, single, and double diffraction are rapidity-gapped forward scattering processes in which zero, one, and two of the initial protons dissociate into jets, respectively, and no central radiation is detected.}\label{fig:diffract}
\end{figure}
 
A $pp$ forward scattering process with no central radiation is \textit{elastic} (and diffractive) if both protons remain intact, \textit{single diffractive} (SD) if one proton dissociates, and \textit{double diffractive} (DD) if both protons dissociate into jets. 
A process is \textit{central diffractive} (CD) if central particles are produced that are rapidity-gapped from the two forward hadronic regions \cite{LHCForwardPhysicsWorkingGroup:2016ote}; CD is about 1\% of the total cross section \cite{ATLAS:2012djz}. Experimentally we can only detect particles above a minimum energy threshold $E_{\rm cut}$, so one does not know if 
a rapidity gap is truly empty or filled with invisible soft radiation. The latter is an unavoidable background for DD where nonsinglet exchange is allowed and is called \textit{quasi-diffraction} \cite{Lee:2025fml}.

The factorization of $pp$ diffraction has long proven elusive. The diffractive PDFs (dPDFs) that appear in small-$|t|$ $ep$ diffraction \cite{Berera:1994xh, Berera:1995fj,Collins:1997sr} and the Ingelman-Schlein model \cite{Ingelman:1984ns} largely fail to describe $pp$ diffractive data \cite{UA8:1992iws, CDF:1993slg, CDF:2000rua}. The literature terms this \textit{factorization breaking}, and often inserts a new \textit{gap survival probability} term \cite{Dokshitzer:1987nc, Bjorken:1991xr} to model the difference \cite{LHCForwardPhysicsWorkingGroup:2016ote}. A second approach is the Good-Walker formalism describing fluctuations in the proton's structure \cite{Good:1960ba,Mantysaari:2020axf}, which is often used to model low-mass SD and DD.  A third approach is the old formulation of \textit{Regge theory} \cite{Regge:1959mz, Collins:1977jy, Donnachie:2002en, Gribov:2003nw}, which expresses scattering amplitudes as a sum $A \propto \sum_n c_n s^{\alpha_n(t)}$, with coefficients $c_n$ and \textit{Regge trajectories} $\alpha(t) = \alpha_0 + \alpha_1 t$. This approach fails to describe data, and so-called triple Pomeron/Reggeon formulas are typically employed in global fits \cite{Caneschi:1969ck, Kancheli:1970gt, Mueller:1971ez}.  Modern Regge methods have superseded old Regge theory, but so far are only applied to amplitudes, not diffractive cross-sections \cite{DelDuca:2011wkl, DelDuca:2011ae, DelDuca:2013ara, Caron-Huot:2013fea, DelDuca:2014cya, Fadin:2016wso, Caron-Huot:2017fxr, Caron-Huot:2017zfo, Fadin:2017nka, Falcioni:2021dgr, Falcioni:2021buo, Falcioni:2020lvv, DelDuca:2021vjq, Caola:2021izf}.   The S-matrix bootstrap can also constrain Regge limits~\cite{Correia:2025uvc,deRham:2025vaq} which our findings may provide useful input for.
 
Recently, a new first-principles approach to forward scattering was developed: a top-down effective field theory (EFT) of QCD called soft collinear effective theory (SCET) \cite{Bauer:2000ew, Bauer:2000yr, Bauer:2001ct, Bauer:2001yt} with Glauber operators \cite{Rothstein:2016bsq}. SCET was used to factorize the Regge dynamics in inclusive DIS~\cite{Neill:2023jcd}, $ep$ diffraction \cite{Lee:2025fml}, and 2-to-2 forward scattering amplitudes \cite{Moult:2022lfy, Gao:2024qsg, Gao:2024fyz}, and we use it here to factorize elastic, single, and double diffraction. 
\\[-8 pt]

\noindent\textbf{\textit{Kinematics.}} Let us consider $pp \to X_1 X_2$, where the $X_i$ can either be intact protons or jets separated by a rapidity gap, as in fig.~\ref{fig:diffract}. The incoming protons carry momentum $p_i$ and exchange momentum $\tau$, forming outgoing states $X_i$ with momenta $p_{X1}=p_1-\tau$ and $p_{X2}=p_2 + \tau$. We take $p_1^2 = p_2^2 = m_p^2 \approx 0$, and thus have up to four linearly independent scales:
\begin{align}\label{eq:mass-scales}
	&s = (p_1 + p_2)^2\,,
	&& t = \tau^2 \,,
	\nn\\
	&M_1^2 = (p_1-\tau)^2\,,
	&&M_2^2 = (p_2 + \tau)^2\,,
\end{align}
with fewer invariants when $M_i^2 =m_p^2 \approx 0$. The literature often expresses $\xi_i  = M_i^2/s$.

We work in the center of mass frame, defined as 
\begin{align}
	&p_1^\mu = \frac{m_p^2}{P} \frac{\nbar^\mu}{2} + P \frac{n^\mu}{2}\,,
	&&p_2^\mu = P \frac{\nbar^\mu}{2} + \frac{m_p^2}{P} \frac{n^\mu}{2}\,,
\end{align}
where $\nbar^\mu = (2,0,0)$ and $n^\mu = (0,2,0)$ form a basis for lightcone coordinates $(+,-,|\!\!\perp\!\!|) = (n\cdot p, \nbar \cdot p, |p_\perp|)$. 
Here $P=\sqrt{s}$, dropping terms of order $\cO(m_p^2/s)$. We can now decompose the remaining momenta as 
\begin{align}\label{eq:com2}
	\tau &= \sqrt{s}\left(\frac{t-M_1^2}{s}, \frac{M_2^2-t}{s},\frac{|\tau_\perp|}{\sqrt{s}} \right)\,,
	\nonumber\\
	p_{X1} &= \sqrt{s} \left(\frac{M_1^2-t}{s}, 1 - \frac{M_2^2-t}{s},\frac{|\tau_\perp|}{\sqrt{s}}  \right)\,,
	\nonumber\\
	p_{X2} &= \sqrt{s} \left(1+\frac{t-M_1^2}{s}, \frac{M_2^2-t}{s},\frac{|\tau_\perp|}{\sqrt{s}} \right)\,,
\end{align}
where from $\tau^2 = t$ we have that
\begin{align}
	|\tau_\perp| = \sqrt{-t - \frac{(M_1^2 -t)(M_2^2-t)}{s}}\,.
\end{align}
Diffraction constrains us to forward kinematics:
\begin{align}\label{eq:forward-constraint}
	&|t| \ll s\,,
	&& M_i^2 \ll s  \,.
\end{align}
We also require a large gap between the rapidities $\eta_k =\frac{1}{2} \log (k^-/k^+)$ of the final hadronic states:
\begin{align}\label{eq:gap-constraint}
	p_{X1}^-/p_{X1}^+ \gg p_{X2}^-/p_{X2}^+\,.
\end{align}
Eq.~\eqref{eq:gap-constraint} is trivially satisfied given eq.~\eqref{eq:forward-constraint}. Typically, experimentalists take the gap to be $\Delta\eta \gtrsim 2$-3 \cite{ATLAS:2012djz,CMS:2015inp, ATLAS:2019asg}. 
\\[-8 pt]

\begin{figure}
	\includegraphics[width = 1.5 in]{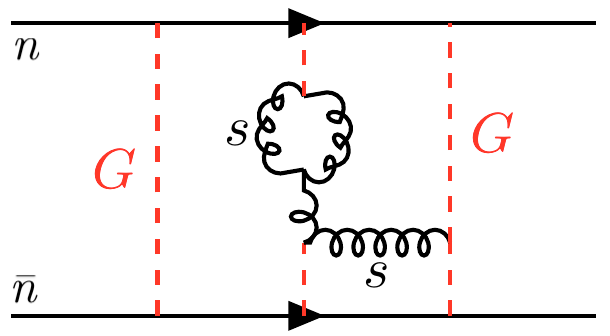}
	\raisebox{0.3 mm}{\includegraphics[width = 1.5 in]{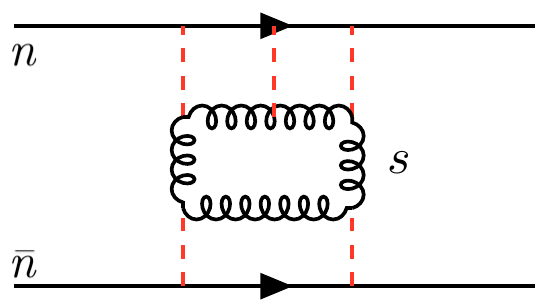}}
	\caption{Diffraction is mediated by Glauber gluon ($G$) exchanges that can be simple, dressed by soft fermion/gluon loops, or interact via soft exchanges; see eq.~\eqref{eq:nsn-ops}. The protons need not emit equal numbers of Glaubers.}\label{fig:soft-loops}
\end{figure}

\noindent\textbf{\textit{Power counting.}} 
We can understand the effect of eqs.~\eqref{eq:forward-constraint} and \eqref{eq:gap-constraint} by defining \textit{power counting} parameters
\begin{align}
	&\lambda= \frac{\sqrt{-t}}{\sqrt{s}}\,,
	&& \rho_i = \frac{M_{i}}{\sqrt{-t}}\,,
	&& \lambda_\Lambda = \frac{\Lambda_{\rm QCD}}{\sqrt{-t}}\,.
\end{align}
From eqs. \eqref{eq:forward-constraint} and \eqref{eq:gap-constraint}, we have that
\begin{align}\label{eq:single-constraints}
	&\lambda \ll 1\,,
	&& \lambda\rho_1 \ll 1\,,
	&& \lambda\rho_2 \ll 1\,.
\end{align}
This implies that diffraction requires $\lambda \ll 1$, but we can have $\rho_i \ll 1$, $\rho_i \sim 1$, or $\rho_i \gg 1$ and still maintain a rapidity gap, so long as $\lambda \rho_i$ remains small.  $\lambda_\Lambda$ determines whether or not the scale $t$ is perturbative. 

We can express eq.~\eqref{eq:com2} in terms of power counting parameters. For the basic case $\lambda\ll 1$ and $\rho_i\sim1$, we have
\begin{align}\label{eq:power-count-com2}
	p_1 &\sim \sqrt{s}(0,1,0) 
	&p_2 &\sim \sqrt{s}(1,0,0) 
	\nonumber\\
	p_{X1} &\sim \sqrt{s} \big(\lambda^2 ,  1,\lambda   \big)
	&p_{X2} &\sim \sqrt{s} \big(1, \lambda^2,\lambda \big)\,.
	\nonumber\\
	\tau &\sim \sqrt{s}\big(\lambda^2, \lambda^2,\lambda \big)
\end{align}
Here, $p_1$ and $p_{X1}$ have $\nbar$-collinear scaling, $\tau$ is Glauber, and $p_2$ and $p_{X2}$ are $n$-collinear. Thus, SCET is the appropriate EFT for factorizing $pp$ diffraction.

We can also predict how much undetected radiation penetrates the gap, noting from the jet mass constraint in eq.~\eqref{eq:forward-constraint} and the power counting in eq.~\eqref{eq:power-count-com2} that gap  modes must be ultrasoft ($us$) \cite{Lee:2025fml}:
\begin{align}\label{eq:gap-radiation}
	p_{us} \sim \sqrt{s} (\lambda^2, \lambda^2, \lambda^2)\,,
\end{align}
using that $p_{us}^2 \sim s\lambda^4$ is significantly smaller than any of the other energy scales $\sim\{s\lambda^0, s\lambda^2\}$ in eq.~\eqref{eq:mass-scales}.
We group the $us$ radiation into a state $Z_{us}$ whose particles can populate both $X_1$ and $X_2$.
\\[-8 pt] 

\noindent\textbf{\textit{SCET operators.}}
In SCET, Glauber operators mediate interactions between soft, $n$, and/or $\nbar$-collinear sectors through the two- and three-sector operators
\begin{align}\label{eq:nsn-ops}
	\cO^{ij}_{ns} &= \mathcal{O}_n^{iB} \frac{1}{\mathcal{P}_\perp^2}\, \mathcal{O}_s^{j_nB}\,,
	\nn\\
	\cO_{ns\nbar}^{ij}&= \cO_n^{iB}\frac{1}{\cP_\perp^2}\cO_s^{BC}\frac{1}{\cP_\perp^2} \cO_{\nbar}^{jC}\,,
\end{align}
whose detailed form can be found in ref.~\cite{Rothstein:2016bsq}.
Here $\cP_\perp$ is the SCET transverse momentum operator, $i,j \in \{q,g\}$, and $B,C$ are octet color indices. Soft ($s$) scaling is $\sim \sqrt s (\lambda,\lambda,\lambda)$.  $\cO_{n,\nbar,s}^{iB}$ are fermion or gluon bilinears connected by color and Dirac structures, e.g.
\begin{align}
	\cO_n^{qB} = \bar{\chi}_n T^B \frac{\slashed\nbar}{2}\chi_n\,,
\end{align}
whereas $\cO_s^{BC}$ is a more complicated product of soft Wilson lines. Fig.~\ref{fig:soft-loops} illustrates some of the types of interactions that these operators describe.
Note that the clean separation of Glauber operators into multiplicative $n$, $s$, and $\nbar$ components will enable us to write down a factorization, grouping each $\cO_{n,\nbar,s}$ into matrix elements with its corresponding states, and pulling out overall Glauber potential factors $\cP_\perp^{-2}$. 

We can describe ultrasoft radiation that penetrates the rapidity gap following ref.~\cite{Lee:2025fml}. We decouple $us$ from $n$-collinear radiation using a BPS field redefinition \cite{Bauer:2001ct}, as they only interact eikonally:
\begin{align}\label{eq:bps}
	&\xi_{n}(x) \to U_{n}(x) \xi_{n}(x),
	&&\cB_{n\perp}^{a\mu}(x) \to \cU_{n}^{ab}(x) \cB_{n\perp}^{b\mu}(x).
\end{align}
We define the Wilson line operators $\mathcal{U}_n^{BB'} T^{B'} = U_n^\dagger T^B U_n$ in terms of the standard Wilson line
\begin{align}
	&U_n(x) = \bar{P} \exp \big[-ig\int_0^\infty ds\, n\cdot A_{us}^A(ns+x)T^A\big]\,.
\end{align}
Here, $T$ is a color generator in the fundamental representation and $\bar{P}$ indicates anti-path-ordering. The $\nbar$ sector behaves similarly. The Glauber operators become
\begin{align}
	&\cO_{n}^{iA} \to \cU_n^{AB} \cO_n^{iB}\,, 
	&\cO_{\nbar}^{jA} \to \cU_{\nbar}^{AC} \cO_{\nbar}^{jC}\,,
\end{align}
which modifies eq.~\eqref{eq:nsn-ops} as
\begin{align}\label{eq:glaub-ops}
	&\cO_{ns}^{ij} \to {\cU}_n^{AB} \cO_n^{iB}\frac{1}{\cP_\perp^2}\cO_s^{j_{n}A}
    \,,
	\nn \\
	&\cO_{ns\nbar}^{ij} \to {\cU}_n^{AB} {\cU_{\nbar}}^{CD} \cO_n^{iB}\frac{1}{\cP_\perp^2}\cO_s^{AC}\frac{1}{\cP_\perp^2} \cO_{\nbar}^{jD}
    \,.
\end{align}
The soft sector is unaffected by the transform in eq.~\eqref{eq:bps}. 
\\[-8 pt]

\noindent\textbf{\textit{Factorization of diffraction.}} Using SCET, we can factorize elastic, single, and double (quasi-)diffraction simultaneously. We start with the CM frame differential cross section for double $pp \to X_1 X_2$ scattering in QCD:
\begin{align}\label{eq:full-xsec}
		&\frac{d\sigma^{\rm DD}}{dM_1^2 dM_2^2 dt} = \frac{1}{2s} \sumint_{X_1,X_2} \prod_{i=1}^2\delta(M_{i}^2 - p_{Xi}^2)
		\\
  &\qquad \times (2\pi)^4 \delta^{(4)}(p_1+p_2 - p_{X1} - p_{X2} ) 
   \nn\\
		&\qquad\times \delta\big((p_{X_2} - p_2)^2 - t\big) 
  \langle p_1 p_2 | X_1 X_2 \rangle_A
  \langle X_1 X_2 |p_1 p_2\rangle_A \,. 
		\nn
\end{align}
Here, the $\sumint$ sums over the phase space of all final-state particles, and as usual $\langle p_1 p_2 | X_1 X_2 \rangle = (2\pi)^4 \delta^{(4)}(\sum k_i) 
\langle p_1 p_2 | X_1 X_2 \rangle_A$. 
We drop the subscript $A$ below.  Elastic and single diffraction are special cases of this cross section, with $X_i = p$.

\begin{figure}
	\includegraphics[width =0.45\textwidth]{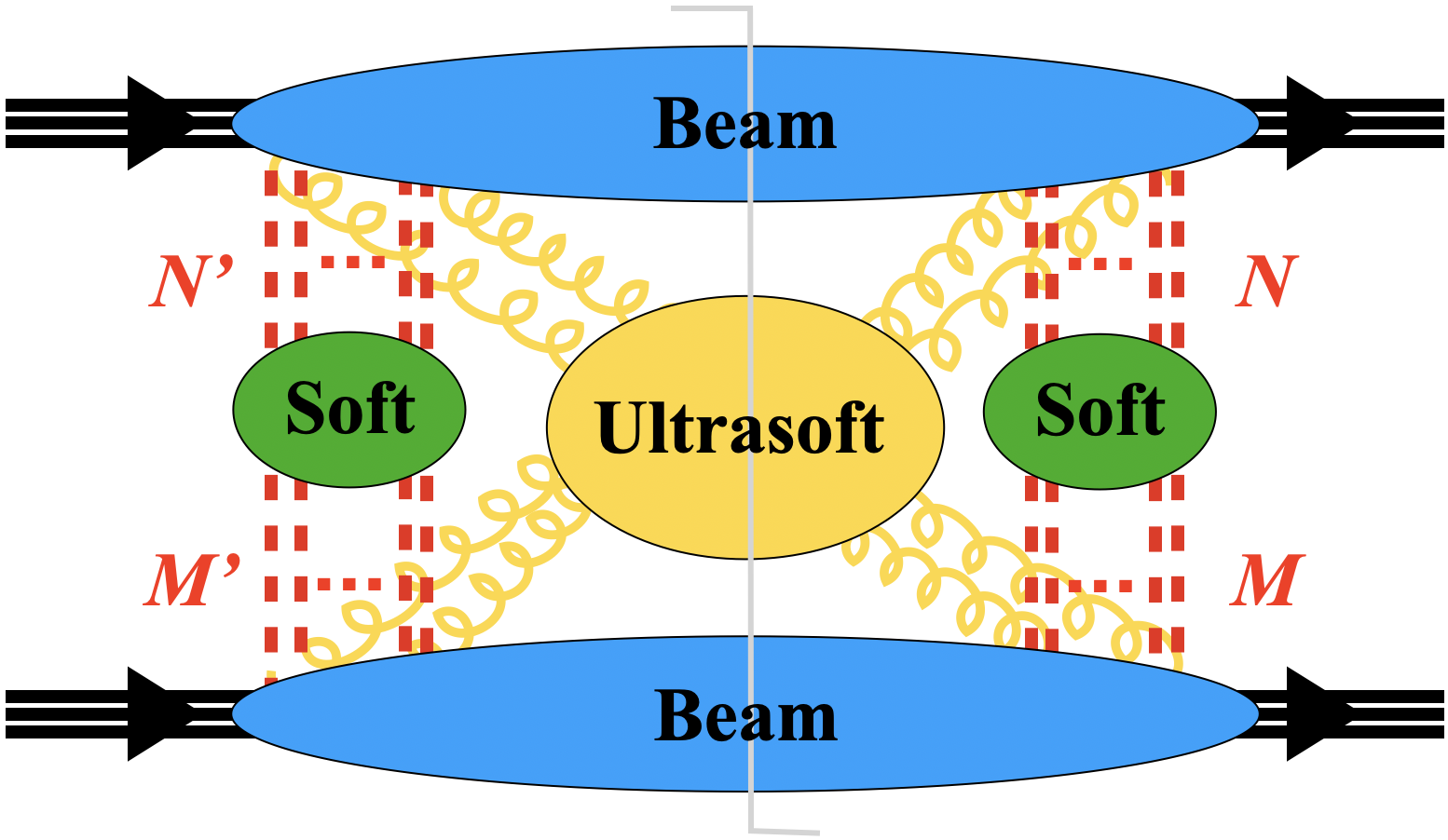}
	\caption{Factorization of double diffraction into matrix elements connected by $M+N$ Glauber exchanges, eq.~\eqref{eq:fact}. Undetected ultrasoft radiation can penetrate the gap under color-nonsinglet exchange.
}\label{fig:fact}
\end{figure}

Using SCET, we separate the states in eq.~\eqref{eq:full-xsec} into soft, collinear, and ultrasoft components:
\begin{align}\label{eq:state-factor}
	&\langle p_1 p_2 | = \langle p_1^n | \langle p_2^{\nbar} | \langle 0_s | \langle 0_{us}|
	\,,\nn\\
	&| X_1 X_2 \rangle = | X_1^n \rangle | X_2^{\nbar} \rangle |0_s \rangle |Z_{us} \rangle \,,
\end{align} 
where $X_1^n$ and $X_2^{\nbar}$ denote collinear radiation in these $X_i$'s.
Just as in $ep$ diffraction \cite{Lee:2025fml} and forward $2\to 2$ scattering amplitudes \cite{Moult:2022lfy, Gao:2024qsg, Gao:2024fyz}, these modes interact with one another through the exchange of an infinite tower of Glauber operators $\cO_{ns\nbar}, \cO_{ns},\cO_{\nbar s}$. Expanding in the Glauber exchanges, we can group operators with their corresponding states in eq.~\eqref{eq:state-factor}, giving a factorization formula for double diffraction
\begin{align}\label{eq:fact}
	&\frac{d\sigma^{\rm DD}}{dM_1^2 dM_2^2 dt} = \frac{(\bar{n}\cdot p_1)^2({n}\cdot p_2)^2}{2s\,}\!\!
	\sum_{\substack{N,N'=1\\M,M'=1}}^{\infty} \sum_{\{R_X\}} \IInt_{NM}^\perp \IInt_{N'M'}^\perp 
	 \nn \\
	&\!\times \delta(t-\tau^2_\perp) \int dp_g^+ dp_g^-
	B_{NN'}^{R_U^{N}R_U^{N'}}\!\!\big( M_1^2 - p_1^- p_g^+ ,\{\tau_{i\perp}^U,\tau'^{U}_{j\perp}\}, t\big) \,
	\nn\\
	&\!\times B_{MM'}^{R_L^M \! R_L^{M'}}\!\!\big( M_2^2 - p_2^+ p_g^-,\{\tau_{k\perp}^L,\tau_{l\perp}'^{L}\}, t\big)
	\nn\\
	&\!\times  S_{NM}^{R_S^N\! R_{\bar S}^M}\!\!\big(\{\tau_{i\perp}^U,\tau^L_{k\perp}\}, t\big)\, \,
	S_{N'M'}^{R_S^{N'}\!R_{\bar S}^{M'} }\!\!\!\big(\{\tau_{j\perp}'^U,\tau'^L_{l\perp}\}, t\big) \,
	\nn\\
	&\!\times U_{NN'MM'}^{R_U^N R_U^{N'}  \!R_L^M R_L^{M'} \! R_S^N R_{\bar S}^M R_S^{N'}\!R_{\bar S}^{M'}}\!\!\big(p_g^+, p_g^-\big)\,,
\end{align} 
as illustrated in fig.~\ref{fig:fact}.
The beam function ($B$), soft functions ($S$), and ultrasoft function ($U$) are matrix elements that we describe in detail below.
The variables $N,M$ indicate how many Glaubers the upper ($^U$) and lower ($^L$) proton emit, carrying momenta $\tau_{i\perp}^U$ and $\tau_{j\perp}^L$ respectively. The presence/absence of a prime ($^\prime$) signifies the left/ right side of the cut. $R_X$ indicates color representation. 

The first beam function in eq.~\eqref{eq:fact} is given by
\begin{align}\label{eq:beam}
	&B_{NN'}^{R^N \!R^{N'}}\!\!\!\big(p^- k^+,\{\tau_{i\perp},\tau'_{j\perp}\}, t\big)
   \!=\! \frac{1}{(\bar{n}\cdot p)^2} \sumint_{X_n} \!\int \!\frac{ dv^-}{2 p^-}e^{\frac{i }{2}v^- k^+}
	\nn\\
	&\!\!\times\langle p |\! \collapsel\!\! \prod_{i=1}^{N\!-\!1}\!\! \cO_n^{A_i}\! \bar\cO_n^{A_N}\!\!\collapser \! P_N^{R^N}\!| \!X_n \rangle\!
	\langle \!X_n | \! P_{N'}^{R^{N'}}\! \!\collapsel\!\!\! \prod_{j=1}^{N'\!-\!1}\!\!\! \cO_n^{A_j}\!\bar \cO_n^{A_{N'}}\!\!\collapser\! | p \rangle.
\end{align}
The formula for the $\nbar$-collinear $B$ is the same, with $n\leftrightarrow \nbar$. 
Here, $\scollapsel ...\scollapser$ indicates \textit{Glauber collapse}: $\tau_{i\perp}$ loop integration takes all the $\cO_n$ to the same $\pm$ coordinates; see Refs.~\cite{Rothstein:2016bsq,Lee:2025fml}.
$P_N^R$ is the projector for an $N$-Glauber state onto color representation $R$ and $\bar \cO_n^A = \int \ddslash^{d-2}\tau_\perp \cO_n^A$; see eq. (4.9) of \cite{Lee:2025fml}. Note that we must project onto same-dimensional representations on both sides of the cut, but these representations can be formed from different numbers of Glaubers $N,N'$ and have different properties like $\cC$-parity. The full expression for $B$ with arguments is identical to the $ep$ beam function in eq. (4.24) of ref.~\cite{Lee:2025fml}.

The ultrasoft function takes the form 
\begin{align}\label{eq:us}
	&U_{N...M'}^{R_U^N ... R_{\bar S}^{M'}}\!\!\big(p_g^+, p_g^-\big) 
	\!\!=\!\! \sumint_{Z_{us}} \!\delta\!\left(p_g^+-p_{us}^{1,+}\right) \delta\!\left(p_g^- - p_{us}^{2,-}\right)
	\\
	& 
	\times\langle 0 | P^{R_U^N}_N P^{R_L^M}_M \prod_{i,j=1}^{N,M} \cU_n^{B^U_i A^U_i} \cU_{\nbar}^{B^L_j A^L_{j}} P^{R_S^N}_N P^{R_{\bar S}^M}_{M}| Z_{us}\rangle 
	\nn\\
	&\times \langle Z_{us} | P^{R_U^{N'}}_{N'} \!P^{R_L^{M'}}_{M'}\! \prod_{k,l=1}^{N',M'} \cU_n^{B^{\prime U}_k \!A^{\prime U}_{k}} \cU_{\nbar}^{B^{\prime L}_l \!A^{\prime L}_l} P^{R_S^{N'}}_{N'} \!P^{R_{\bar S}^{M'}}_{M'}\!| 0\rangle\,,
	\nn
\end{align}
where $p_{us}^{i,\pm}$ is a $us$ radiation component in hemisphere $i$ that contributes to that $i$'s jet mass.
Ultrasoft modes connect beam representations $(R_U, R_L)$ characterized by color indices $A$, to soft representations $(R_S, R_{\bar S})$ with color indices $B$. 

The soft function is
\begin{align}\label{eq:soft}
	S_{NM}^{R_S^N R_{\bar S}^M}\!\big(\{\tau_{i\perp}^U,\tau^L_{k\perp}\}, t\big)&=
	\langle 0 | P_N^{R_S^N} \bO_{NM} P_M^{R_{\bar S}^M} | 0\rangle,
\end{align}
and equivalently for the other side of the cut. Here, we use $\bO$ to indicate the intricate Glauber dressings and interactions discussed in fig.~\ref{fig:soft-loops}, which for concision we do not specify explicitly here. Nonetheless, it is straightforward to work out the relevant combination of soft operators and hence soft diagrams order-by-order as needed, see e.g.~\cite{Rothstein:2016bsq,Gao:2024qsg}.

Finally, we have the transverse convolution structure 
\begin{align}\label{eq:conv}
	\IInt_{\substack{NM}}^\perp &=	\frac{(-i)^{N+M}}{N! M!}
	\int \prod_{i,k=1}^{N,M}\frac{\ddslash^{d-2}\tau^U_{i\perp}}{\tau_{i\perp}^{U\,2}}\: 
	\deltaslash^{d-2}\!\big(\textstyle\sum \tau^U_{i\perp} - \tau_{\perp} \big)  \nn \\
	& \times\frac{\ddslash^{d-2}\tau^L_{k\perp}}{\tau_{k\perp}^{L\,2}} \:
	\deltaslash^{d-2}\!\big(\textstyle\sum\tau^L_{k\perp} - \tau_{\perp} \big)  \,.
\end{align}
We have $\tau_\perp \equiv (p_{X_2}^{\bar{n}}-p_2)_\perp =-(p^n_{X_1}-p_1)_\perp$, where the last equality is enforced through the soft function.
We also use $\ddslash^d k = d^dk/(2\pi)^d$ and $\deltaslash^d = (2\pi)^d \delta^d$.
\\[-8 pt]

\begin{figure}
	\includegraphics[width = 2 in]{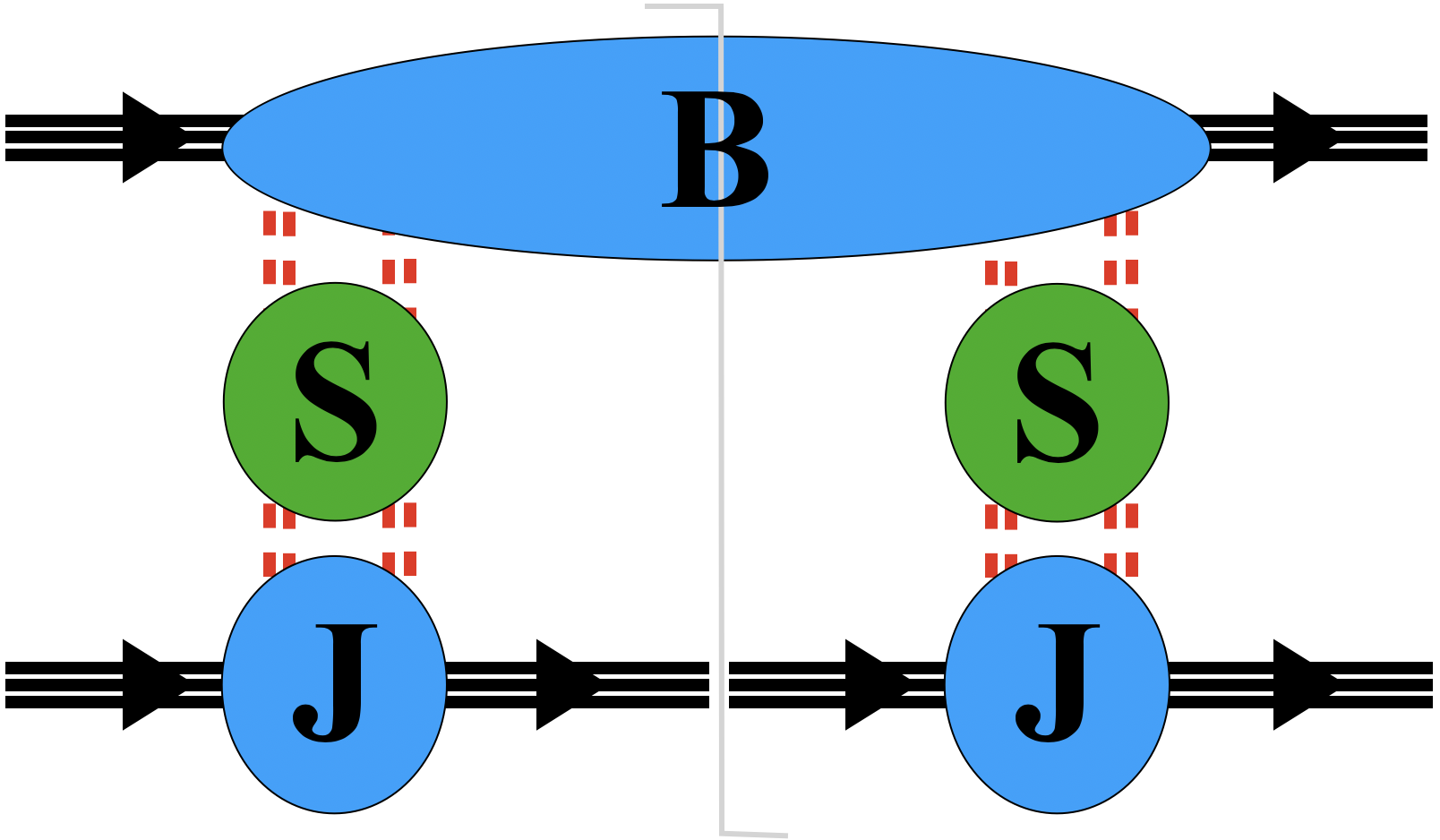}\qquad
	\includegraphics[width = 1 in]{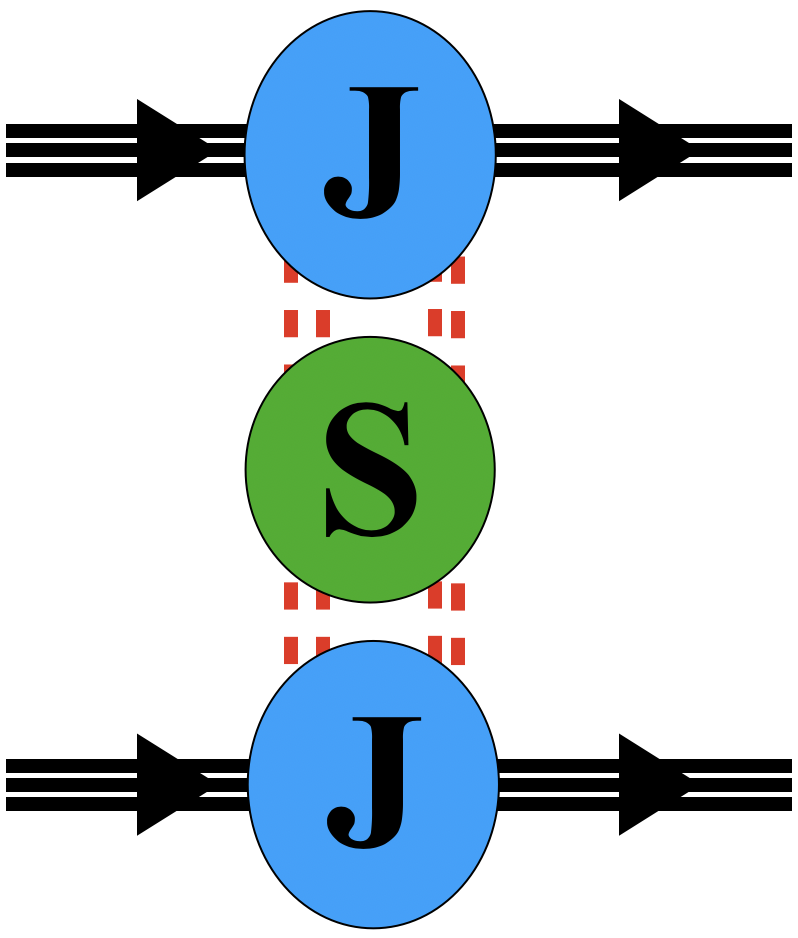}
	\caption{Fig.~\ref{fig:fact} simplifies when one or both protons remain intact. (Left) Single diffraction, eq.~\eqref{eq:sd-fact}. (Right) Elastic diffraction amplitude, eq.~\eqref{eq:ed-fact}.}\label{fig:sd-ed-fact}
\end{figure}

\noindent\textbf{\textit{Single diffraction (SD).}} Eq.~\eqref{eq:fact} is the general form for DD but simplifies for SD, which has an intact forward proton and thus must be mediated by a color-singlet $t$-channel exchange. Just as in $ep$ diffraction \cite{Lee:2025fml}, the Wilson lines in eq.~\eqref{eq:us} collapse under color-singlet projectors $P_N^R$, $U$ becomes a $\delta$-function in color and momenta, no radiation penetrates the gap, and eq.~\eqref{eq:fact} reduces to a simpler factorization structure; see fig.~\ref{fig:sd-ed-fact}:
\begin{align}\label{eq:sd-fact}
	&\frac{d\sigma^{\rm SD}}{dM_1^2 dt} \!=\! \frac{(\bar{n}\!\cdot\! p_1\,{n}\!\cdot\! p_2)^2}{2s\,}\!\!\!\!
	\sum_{\substack{N,N'=1\\M,M'=1}}^{\infty} 
	\sum_{\{R_X=1\}}\!
	\IInt_{NM}^\perp\IInt_{N'M'}^\perp 
	\\
	&\!\times \! B_{NN'}^{R_U^{N}R_U^{N'}}\!\!\!\big( {M_1^2} ,\{\tau_{i\perp}^U,\tau'^{U}_{j\perp}\}, t\big) \,
	J_{M}^{R_L^M}\!\!\big({\{\tau_{k\perp}^L\}, t}\big)	\,
	J_{M'}^{R_L^{M'}}\!\!\!\big(	{\{\tau'^L_{l\perp}\}, t}\big) \, 
	\, \nn \\
	& \! \times
  S_{NM}^{{R_U^N R_L^M}}\!\!\big(\{\tau_{i\perp}^U,\tau^L_{k\perp}\}, t\big)\,
  S_{N'M'}^{R_{U}^{N'}\!R_{L}^{M'} }\!\!\!\big(\{\tau_{j\perp}'^U,\tau'^L_{l\perp}\}, t\big)\,
  \delta(t -\tau_\perp^2) \,.
  \nn
\end{align}
This is less differential than eq.~\eqref{eq:fact}, as $M_2^2=m_p^2 \approx 0$. For the intact proton, we take $\sumint_{X_n} |X_n \rangle\langle X_n| \to |p\rangle\langle p|$ in eq.~\eqref{eq:beam} (see \S 7.4 of ref.~\cite{Lee:2025fml}), which splits $B$ into two amplitudes $J$, via $\lim_{\delta\to 0}\int_{m_p^2-\delta}^{m_p^2+\delta}\! dM_2^2\, B = (J)^2$ :
\begin{align}\label{eq:jet}
	J_{M}^{R_L^M} = \frac{1}{({n}\cdot p_2)}\langle p_2 | \collapsel\!\! \prod_{i=1}^{M-1}\!\! \cO_n^{A^L_i}\bar\cO_n^{A^L_M}\!\!\collapser P_M^{R_L^M}|p_{X2}^{\nbar} \rangle\,.
\end{align}
We call $J$ radiative jet functions, as in Refs.~\cite{Moult:2022lfy, Gao:2024fyz, Gao:2024qsg}. 
\\[-8 pt]

\noindent\textbf{\textit{Elastic diffraction (ED).}} ED is in turn a special case of eq.~\eqref{eq:sd-fact} in which both protons remain intact. Because this process is fully exclusive, the factorization simplifies to occur at the amplitude level; see fig.~\ref{fig:sd-ed-fact}:
\begin{align}\label{eq:ed-fact}
	&\frac{d\sigma^{\rm ED}}{dt} \!=\! \frac{(\bar{n}\cdot p_1\,{n}\cdot p_2)^2}{2s\,}
  \bigg|\!
	\sum_{\substack{N,M=1}}^{\infty}\!
	\sum_{\{R_X\}} 
	\IInt_{\substack{NM}}^\perp \!\!
	J_{N}^{R_U^N}\!\!\big({\{\tau_{i\perp}^U\}, t}\big) 
	 \nn \\
	&\!\times J_{M}^{R_L^M}\!\!\big({\{\tau_{k\perp}^L\}, t}\big) \,  
	S_{NM}^{R_U^N R_L^M}\!\!\big(\{\tau_{i\perp}^U,\tau^L_{k\perp}\}, t\big)\bigg|^2\!
  \delta\!\left(t-{\tau_\perp^2}\right)
  \,,
\end{align}
where $M_1^2=M_2^2=m_p^2 \approx 0$. The amplitude appearing in this formula corresponds to those in Refs.~\cite{Moult:2022lfy, Gao:2024qsg, Gao:2024fyz}.
\\[-8 pt]

\noindent\textbf{\textit{Non-universality of $ep$ versus $pp$.}} 
To isolate and test for the universality of nonperturbative physics we must consider which functions can depend on the nonperturbative scale $\Lambda_{\rm QCD}$. After the $\lambda\ll 1$ expansion, this includes $B$, $J$, and $S$. 
If $\lambda_\Lambda\sim 1$, then all three functions are fully nonperturbative. In this case,
the functions $B$ and $J$ are universal between $ep$ and $pp$, but the soft functions $S$ are non-universal. 
In particular, in $ep$ diffraction, the soft function $S$ involves an electromagnetic current and Glauber operators $\mathcal{O}_s^{j_n}$, whereas in $pp$ diffraction, there is no electromagnetic current and a larger set of Glauber operators appears; namely, $\mathcal{O}_s^{j_{\nbar}}$, $\mathcal{O}_s^{j_n}$ and $\mathcal{O}_s$. This non-universality of $S$ implies that the nonperturbative hadronic functions for $ep$ and $pp$ diffraction differ. 
For the alternate case $\lambda_\Lambda\ll 1$, we can further factorize $S$ and $B$, which 
opens up a possibility for identifying universal hadronic functions, as we discuss below.

As shown in ref.~\cite{Lee:2025fml}, the diffractive PDF (dPDF) for $ep$ is given by a combination of $B$ and a soft-collinear function factorized from the $ep$ soft function in the limit $-t \ll Q^2$. Given the lack of a hard scale like $Q$ in the $pp$ case, there is no analogous factorization of the soft function $S$, and hence there is no sense in which the same dPDFs appear. 
This is true for both $\lambda_\Lambda\sim 1$ and $\lambda_\Lambda \ll 1$. 

Furthermore, eq.~\eqref{eq:fact} does not resemble the Ingelman-Schlein model. $B$ and $S$ are connected by an infinite number of $\perp$ convolutions rather than a single longitudinal exchange. 
Further, eq.~\eqref{eq:fact} has no direct analogue of a rapidity gap survival probability. 
\\[-8 pt]

\noindent\textbf{\textit{Universality of RG evolution.}}
Let us start by considering the color-singlet channel, which includes ED, SD, and the singlet channel of DD. 
By determining the rapidity renormalization group (RRG) equations~\cite{Chiu:2011qc,Chiu:2012ir} we can sum large logarithms $\ln(-t/s)$.   
Renormalized operators dependent on a rapidity scale $\nu$ relate to bare operators dependent on a rapidity regulator $\eta$ as:
\begin{align}
	&\fat{B}(\eta) = \fat{B}(\nu) \cdot \fat{Z}_{\fat{B}}(\eta,\nu)
	 \nn\\
	&\fat{S}(\eta) = \fat{Z}_{\fat{S}}(\eta,\nu) \cdot \fat{S}(\nu) \cdot \fat{Z}_{\fat{S}}^T(\eta,\nu)  
	\\
	&\fat{J}(\eta) = \fat{J}(\nu) \cdot \fat{Z}_{\fat{J}}(\eta,\nu)\,.
	\nn
\end{align}
Here, bold notation indicates an infinite matrix characterized by number of Glaubers, color, and $\tau_{i\perp}$ dependence as in Refs.~\cite{Gao:2024fyz,Gao:2024qsg}. The operation $\cdot$ encodes contraction of color indices and $\tau_{i\perp}$ integrations.  
More explicitly:
\begin{align}\label{eq:RG_Jet}
	&\left(\fat{J}\cdot\fat{Z}_{\fat{J}}\right)^{R_U^N}_N\!\big(\{\tau_{i\perp}^U\},\eta)
	= \sum_{K=1}^{\infty} \sum_{R}\int_\perp \prod_{k=1}^K\frac{\ddslash^{d-2}\tau_{k\perp}}{K!\,\tau_{k\perp}^2}
	 \nn \\
	&\quad \times \deltaslash^{d-2}\!\big(\textstyle\sum_k \tau_{k\perp} - \tau_\perp\big) \, J^{R^K}_K\!\big(\{\tau_{k\perp}\},\nu\big) \nn \\
	&\quad \times Z_{J;K,N}^{R^K ; R^N_U}\!\big(\{\tau_{k\perp}\},\{\tau_{i\perp}^U\},\nu,\eta\big).
\end{align}
	
To analyze universality it is easiest to start with ED, which has an amplitude-level RRG like in Refs.~\cite{Gao:2024fyz,Gao:2024qsg}. From RRG consistency, we have 
\begin{align}\label{eq:rrge-jet-soft}
	\fat{Z}_{\fat{S}} = \fat{Z}_{\fat{J}}^{-1}
	\,.
\end{align}
The soft functions are universal across ED, SD, and singlet DD, so RRG consistency then implies 
\begin{align}
	\fat{Z}_{\fat{B}} = \fat{Z}_{\fat{S}}^{-1} \fat{Z}_{\fat{S}}^{\prime-1}
   \,. 
\end{align}
As in eq.~\eqref{eq:fact}, a prime ($'$) indicates the left side of the cut, as $S$ is an amplitude-level object. In contrast, for $B$, color, transverse momentum, and the number of Glaubers  occur on each side of the cut independently.
Defining the anomalous dimension matrix:
\begin{align}
	\fat{\Gamma} = -(\nu \partial_\nu \fat{Z}_{\fat{J}})\cdot \fat{Z}_{\fat{J}}^{-1},
\end{align}
the RRG equations (RRGEs) takes the form
\begin{align}\label{eq:rrge}
	\nu \partial_\nu \fat{J}(\nu) &= \fat{J} \cdot \fat{\Gamma} 
 \nn \\
	\nu \partial_\nu \fat{S}(\nu) \,,
    &= - \fat{\Gamma} \cdot \fat{S} - \fat{S} \cdot \fat{\Gamma}^T 
  \nn \,, \\
	\nu \partial_\nu \fat{B}(\nu) &= \fat{B} \cdot \left( \fat{\Gamma} \mathbb{1}' + \mathbb{1} \fat{\Gamma}'\right)
  \,.
\end{align}
Thus, we can analyze the RRGE at the amplitude level. 

Refs.~\cite{Gao:2024fyz,Gao:2024qsg} established a method to access $\fat\Gamma$  
for arbitrary color channels, and our analysis shows that we can directly apply these results to ED, SD, and color-singlet DD. 
Note that only a small subset of color-projected anomalous dimensions $\gamma^R_{(N,M)}$ for $M,N < 4$ have been computed thus far.

DD also has nonsinglet quasi-diffractive contributions with a nontrivial $U$, which leads to additional running in the virtuality scale $\mu$. 
(The rapidity evolution of $S$ and $B$ still cancel one another at the amplitude level.)
This quasi-diffractive $\mu$-evolution will lead to exponential Sudakov suppression. 
Ref.~\cite{Lee:2025fml} studied analogous $\mu$-evolution in $ep$ diffraction at leading logarithmic order, 
and found that despite suppression from $U$,
quasi-diffractive contributions may still be sizable in parts of kinematic phase space.
\\[-8pt]

\noindent\textbf{\textit{Ingredients for predicting $pp$ diffraction.}}
Due to the complexity of our $pp$ diffractive factorization, a deeper understanding of its ingredients is necessary to make phenomenological predictions. 
In particular, eqs.~\eqref{eq:fact} and ~\eqref{eq:rrge} involve infinitely large matrices in color space and in the number of Glauber exchanges, which would only be tractable if we could identify an additional organizing principle.
It is useful to start with color-singlet exchange and consider two cases: $\lambda_\Lambda \ll 1$ and $\lambda_\Lambda \sim 1$. 

For $\lambda_\Lambda \ll 1$, we will need to consider a further factorization of the $S$ and $B$ functions, if we wish to separate perturbative ingredients at $\mu^2\sim -t$ from the intrinsically nonperturbative functions: $S = \hat S \otimes S_{\rm NP}$ and $B = \hat B \otimes B_{\rm NP}$.  This would distinguish perturbative Glauber exchanges with 
$\tau_{i\perp}^2 \gg \LQCD^2$ from nonperturbative Glauber exchanges with 
$\tau_{j\perp}^2 \sim \LQCD^2$.
The Glauber number and color indices carried by $S$ and $\hat S\otimes S_{\rm NP}$ will match, but could be simpler on the individual $\hat S$ or $S_{\rm NP}$ (and similarly for the $B$'s).
This refactorization would likely also induce new invariant mass RG evolution equations. 
Similarly, each anomalous dimension could also have perturbative and nonperturbative contributions $\gamma = \hat \gamma \otimes \gamma_{\rm NP}$.

The full $pp$ factorization formula presented in eq.~\eqref{eq:fact} does not lead to immediate predictions.
$B$ has important nonperturbative contributions even for $\lambda_\Lambda \ll 1$. We cannot use lattice QCD to access $B$, as there is currently no lattice formalism for Glauber operators or the small-$x$ regime. Furthermore,
due to the infinite Glauber number and color spaces involved, 
we do not necessarily have perturbative predictive power over $S$ or $\gamma$.

Several possible simplifications could make eq.~\eqref{eq:fact} more predictive:
\begin{itemize}[topsep = 2 pt,itemsep = 1 pt]
 \item If we could truncate the number of Glaubers relevant for the anomalous dimension, i.e. $N\le K$. This could result from, e.g., nonperturbative exchanges being unable to initiate new color channels or change the rapidity anomalous dimension.
\item If the color-singlet anomalous dimension for different channels $R$,$R'$ were universal for different numbers of Glauber exchanges $\gamma_{(N,M)}^{R} = \gamma_{(N',M')}^{R'}$.
\item If the anomalous dimension were to have a simple separation of perturbative and nonperturbative contributions. For example, transverse momentum distributions exhibit an additive $\gamma = \hat\gamma + \gamma_{\rm NP}$ \cite{Boussarie:2023izj}.
\item  If $S_{\rm NP}$ and $B_{\rm NP}$ were to have a much simpler dependence on the number and color of the Glauber exchanges than $S$ and $B$. 
\end{itemize}
Some of these features have been observed in $ep$ diffraction~\cite{Lee:2025fml}. At lowest order in $\alpha_s$, 
adding any number of nonperturbative Glauber exchanges does not change the octet color channel of the $ep$ soft function,
enabling perturbative predictability with a modified nonperturbative beam function.  Additional color channels, like the color singlet, require two or more perturbative Glauber exchanges or perhaps perturbative $\alpha_s$ corrections to the soft function. This provides hope for a more general truncation of color channels.

Future calculations and refactorizations for $B$ and $S$ will enable us to rule these possibilities in or out for $pp$ diffraction, and thus 
access the full implications of the Regge factorization in eq.~\eqref{eq:fact}. 
This could result from identifying symmetries, exploiting the large $N_c$ limit, or determining organizational principles that emerge from perturbation theory. 
Various outcomes are possible:
\begin{itemize}[topsep = 2 pt,itemsep = 1 pt]
 \item If the anomalous dimension were to exhibit truncations or universality, we could predict its $s$-dependence
 \item If $S_{\rm NP}$ and $B_{\rm NP}$ were simple enough, they could
 be exploited as universal objects for $pp$ diffraction like PDFs. E.g., if they were to involve proton wavefunctions that are independent of the number of Glaubers and color.
 \item If we could factorize NP effects from the $ep$ and $pp$ soft functions, and this results in non-universal perturbative components $\hat S^{ep}\ne \hat S^{pp}$, but the same nonperturbative structure $S_{\rm NP}^{ep}=S_{\rm NP}^{pp}$, then it remains an open possibility that universal 
hadronic functions could be identified for $ep$ and $pp$. 
\end{itemize}  
Our predictive power over target cross sections like $\sigma^{\rm tot}$ and $d\sigma/d|t|$ will therefore depend on the results of future work. 
This could, for example, help characterize the dip-bump region near $|t|\sim 0.5$ GeV$^2$ and the Coulomb-Nuclear Interference region near $|t| \sim 5 \times 10^{-4}$ GeV$^2$ \cite{TOTEM:2016lxj, TOTEM:2017sdy}.
It would also be interesting to explore the impact of nonlinear dynamics \cite{Balitsky:1995ub,Jalilian-Marian:1997jhx, Jalilian-Marian:1997qno, Kovchegov:1999yj, Kovner:2000pt, Iancu:2000hn, Iancu:2001ad, Ferreiro:2001qy} and the saturation transition \cite{Gribov:1983ivg, Mueller:1985wy, Mueller:1989st, Iancu:2003xm, Gelis:2010nm} on $pp$ diffraction observables, using Glauber SCET methods, building on ref.~\cite{Stewart:2023lwz}.

Next consider the case $\lambda_\Lambda\sim 1$, where the full $S$ and $B$ are nonperturbative. Here, there are no truncations in the number of Glauber exchanges,  but there could still possibly be simplifications for the color-singlet anomalous dimensions that enable predictions for the $s$-dependence. 

For the color-nonsinglet channels relevant for double diffraction, it would be useful to explore the same set of possible simplifications in the $\lambda_\Lambda\ll 1$ and $\lambda_\Lambda\sim 1$ cases, in order to characterize the quasi-diffractive background. 
It would also be interesting to carry out resummation, to estimate the Sudakov suppression induced by $U$.
\\[-8 pt]

\noindent\textbf{\textit{Donnachie-Landshoff (D-L) scaling.}}
We can generalize our $pp \to X_1X_2$ factorization to any hadronic collision $h_1 h_2 \to X_1X_2$ by simply replacing the beam function states $|p_1^n\rangle, |p_2^{\nbar}\rangle \to |h_1^n\rangle, |h_2^{\nbar}\rangle $ in eq.~\eqref{eq:beam}. $S$, $U$, and the convolution structure remain unchanged in eq.~\eqref{eq:fact}. 
The universality of $S$ implies that the anomalous dimensions of all colliding hadrons $h_1 h_2$ are identical, though their boundary conditions will differ. This provides partial heuristic evidence for D-L scaling \cite{Donnachie:1992ny}, the observation in the 1990s that for various hadron pairs ($h_1h_2$ = $pp$, $p\pbar$, pions, kaons, ...), experimental total cross section data can be approximately fit to $\sigma^{\rm tot} \approx X_{h1,h2}\, s^{0.0808} + Y_{h1,h2}\, s^{-0.4525}$ for constants $X$ and $Y$ that depend on the colliding particles. 
However, D-L scaling is insufficient to describe later higher-energy collider measurements \cite{TOTEM:2015oop} (and violates the Froissart unitarity bound on cross section growth \cite{Froissart:1961ux}). We hope that further calculations within the SCET formalism will more sharply constrain the functional forms of these cross sections.
\\[-8 pt]

\noindent\textbf{\textit{Odderons.}} The difference between the $pp$ and $p\pbar$ diffractive cross sections is conventionally attributed to the exchange of a three-gluon charge conjugation ($\cC$) odd color-singlet state called the Odderon. Our factorization formula in eq.~\eqref{eq:fact} provides a more rigorous framework for analyzing the difference between $pp$ and $p\pbar$ scattering.  These processes differ only in the state $|p\rangle$ or $|\pbar\rangle$ appearing in \textit{one} beam function,  eq.~\eqref{eq:beam}. Inserting $\hat \cC^\dagger \hat \cC$ gives
\begin{align}
	\langle p | P^R_N... |X\rangle &=  \langle \pbar|  (\hat \cC P^R_N \cdots \hat{\cC}^{\dagger}) |\bar X\rangle
	\nn\\
	&\stackrel{R=1}{=} \pm \langle \pbar|  (P^R_N \cdots) |\bar X\rangle
 \,.
\end{align}
where in the second line, we restrict to the color-singlet case and use that we can always choose our basis of color-singlet projections to have definite $\cC$-parity.
This implies that after the sum over states $\sum_X$, the color-singlet $pp$ and $p\pbar$ cross sections only differ by the terms with a $\cC$-even and a $\cC$-odd projector on opposite sides of the cut ($\cC$-mixed terms), schematically:
\begin{align}\label{eq:odderon}
	d\sigma^{pp}_{R=1} - d\sigma^{p\pbar}_{R=1} &= 2 \,\sum \text{$\cC$-mixed terms}
\end{align}
As gluons are color octets, we need at least three gluons to produce a $\cC$-odd color-singlet exchange. However, higher $\cC$-odd gluon exchanges also contribute, which are not \textit{a priori} suppressed,
as discussed above. Interestingly, from experiment we know that the value of eq.~\eqref{eq:odderon} is small, 
implying that same-sign-$\cC$ exchanges dominate over $\cC$-mixed terms in $pp$ and $p\pbar$ scattering. 
Further Odderon calculations may also provide information on differences in $pp$ \cite{TOTEM:2018psk, TOTEM:2021imi} and $p\pbar$ \cite{D0:2012erd} dip-bump features.
\\[-8 pt]

\noindent\textbf{\textit{Conclusions and outlook.}}
We use SCET to derive factorization formulas for elastic, single, and double $pp$ diffraction. The factorization illuminates why existing models fail to describe data, and why dPDFs determined from $ep$ data fail to describe $pp$ diffraction. Furthermore,  the hadronic functions appearing in $pp$ and $ep$ diffractive factorizations are nonuniversal for $|t|\sim \Lambda_{\rm QCD}^2$, but their rapidity anomalous dimensions are universal. 
For $|t|\gg \Lambda_{\rm QCD}^2$, our analysis leaves open the possibility that the process dependence may be perturbative; however, identifying useful universal hadronic functions still would require miraculous simplifications. 
Additionally, we show that ultrasoft radiation below the detection threshold can penetrate the rapidity gap in double diffraction when the interaction is mediated by color-nonsinglet exchange. 
We discuss the extension of our factorization to other hadron collisions, providing new paths to understanding Donnachie-Landshoff scaling and Odderon measurements. Our results 
lay the groundwork 
for a deeper understanding of diffraction at the LHC.
\\[-8 pt]

\noindent\textbf{\textit{Acknowledgments.}}
We thank Miguel Correia and Matt Schwartz for interesting discussions.
S.T.S. appreciates the hospitality of MIT, Harvard, and Washington University in St. Louis while working on this manuscript.
This work was supported by the U.S. Department of Energy, Office of Science, Office of Nuclear Physics under contracts DE-SC0011090, DE-AC02-06CH11357, and the Quark-Gluon Tomography Topical
Collaboration with award number DE-SC0023646.
S.T.S. was also supported by the Hoffman Distinguished Postdoctoral Fellowship through the LDRD Program of Los Alamos National Laboratory under Project 20240786PRD1. Los Alamos National Laboratory is operated by Triad National Security, LLC, for the National Nuclear Security Administration of the U.S. Department of Energy (Contract Nr. 892332188CNA000001).

\bibliography{diff}
	
\end{document}